\documentclass[%
 reprint,
superscriptaddress,
 amsmath,amssymb,
 aps,
pra,
]{revtex4-2}

\usepackage{graphicx}
\usepackage{bm}
\usepackage{braket}

\usepackage{siunitx}
\usepackage{hyperref}

\begin{document}
\title{
From harmonic to sound-like oscillations in a quantum gas\\
confined in a gravity compensated shell trap
}

\author{Matthieu~Cassus}
\author{Rishabh~Sharma}
\author{Maxime~Pesche}
\author{Laurent~Longchambon}
\author{Thomas~Badr}
\affiliation{Laboratoire de physique des lasers, Université Sorbonne Paris Nord and CNRS UMR 7538, 99 av. J.-B. Clément, 93430 Villetaneuse, France}
\author{Romain~Dubessy}
\affiliation{Aix-Marseille University, CNRS UMR 7345, PIIM, 13397, Marseille, France}
\author{H\'el\`ene~Perrin}
%\email{helene.perrin@univ-paris13.fr}
\affiliation{Laboratoire de physique des lasers, Université Sorbonne Paris Nord and CNRS UMR 7538, 99 av. J.-B. Clément, 93430 Villetaneuse, France}

\begin{abstract}
We study the center of mass oscillations of a quantum gas in a shell shaped trap in the presence of a vertical force opposed to gravity. The measured harmonic frequency at the bottom of the shell is compared with an analytical formula for the trap potential including corrections beyond the rotating wave approximation. When gravity is partially compensated, a quartic correction to the harmonic motion has to be included due to the shell curvature. As gravity is nearly canceled, the quantum gas occupies a large fraction of the lower hemisphere. Driving the center of mass induces internal excitations in the quantum gas, whose time evolution is governed by the presence of sound waves. Relaxation processes induce a strong damping of the center of mass oscillation in this limit.
\end{abstract}

\maketitle

\section{Introduction and main result}
The confinement of quantum gases in shell-shaped traps offers unique opportunity to study the interplay of geometry, topology and dimensionality, motivating a lot of theoretical studies and a few experiments (see~\cite{Lundblad2023,Tononi2024,Dubessy2025} for recent reviews). The shell geometry leads to specific collective modes \cite{Sun2018} and to a total zero vorticity \cite{Turner2010} even in the presence of rotation \cite{Saito2023,White2024}. On a thin shell, the large scale flows of superfluid hydrodynamic present analogies with atmospheric physics  \cite{Tercas2010,Skipp2022,Saito2023}.

The main limitation to realize a filled shell-shaped trap is the effect of gravity that pulls the atoms to the bottom of the shell. In order to mitigate this effect several strategies have been implemented: realize the experiment in micro-gravity \cite{Carollo2022}, use a mixture of quantum gases in an optical trap~\cite{Wolf2022,Jia2022}, or compensate gravity with another force~\cite{Zobay2004,Guo2022}. However a partially filled shell-shaped trap is already interesting to investigate the interplay of rotation and curvature, that will influence the dynamics of quantum vortices on the surface \cite{Dubessy2025}.

\begin{figure}[ht!]
    \centering
    \includegraphics[width=8.6cm]{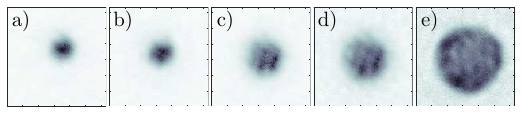}\\
    \includegraphics[width=8.6cm]{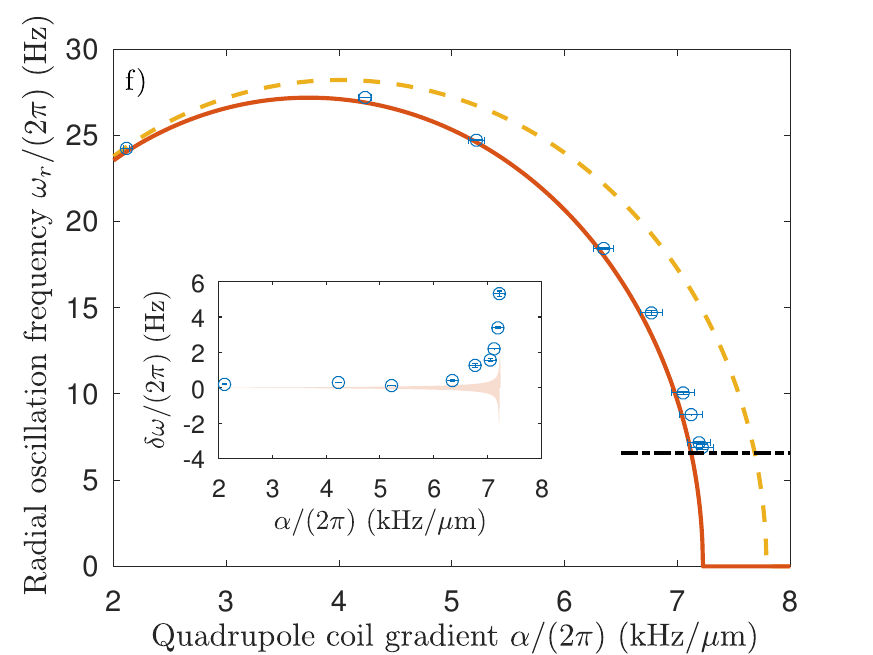}
    \caption{(Color online) a-e) In situ pictures of a quantum gas expanding at the bottom of the shell potential as the quadrupole field gradient is increased $\alpha/(2\pi)=\{2.11,4.23,6.34,6.77,7.19\}\,\SI{}{\kilo\hertz\per\micro\metre}$. The field of view is 120$\times$\SI{120}{\micro\metre}. f) Radial oscillation frequency $\omega_r/(2\pi)$ as deduced from the experimental measurement (blue circles), as a function of the quadrupole field gradient $\alpha/(2\pi)$ which controls the effective gravity, compared to the RWA prediction $\omega_{r,\textrm{RWA}}/(2\pi)$, Eq.~\eqref{eqn:freqRWA} (yellow dashed curve), and to the beyond-RWA formula $\omega_{r,\textrm{Fl}}/(2\pi)$, Eq.~\eqref{eqn:omega_r} (red solid curve), with no adjustable parameter, see text for details. The dashed-dotted horizontal black line gives the expected frequency for an oscillation traveling across the gas at the speed of sound, evaluated for a vanishing harmonic frequency.
Inset: difference $\delta\omega/(2\pi)$ between the measured and expected values computed with the beyond-RWA formula. The shaded area indicates the estimated uncertainty.
    \label{fig:1}
    }
\end{figure}

A radio-frequency (rf) dressed quadrupole trap \cite {Merloti2013} provides such a curved geometry for atoms at the bottom of a shell-shaped trap. As demonstrated in Ref.~\cite{Guo2022}, a vertical gradient in the rf coupling amplitude creates a vertical force pointing against gravity, resulting in an effective reduced gravity. Shrinking the bubble radius enhances this force and the quantum gas progressively fills a larger fraction of the bubble surface, see Fig.~\ref{fig:1}a-e).

In this work, we evidence the associated softening in the center of mass oscillation frequency through the response of the gas to a periodic drive, as illustrated in Fig.~\ref{fig:1}f). We determine precisely the conditions for cancellation of the radial harmonic frequency. Near this point, anharmonicity becomes crucial and the resonant process involves internal excitations, sound waves, that result in a strong damping of the center of mass oscillation.

This paper is organized as follows: in Sec.~\ref{sec:bubble} we explain the mechanism which controls the effective gravity in the bubble. We then detail the experimental protocol and measurements in Sec.~\ref{sec:exp}. In Sec.~\ref{sec:model} we show that an improved modeling of the dressed quadrupole adiabatic potential, including terms beyond the rotating wave approximation, is required to reproduce the results quantitatively. We discuss effects beyond the simple damped harmonic oscillator model in Sec.~\ref{sec:diss} and finally present our conclusions in Sec.~ \ref{sec:conc}.

\section{Gravity compensation in the bubble trap}
\label{sec:bubble}
The curved potential that is used in this paper is obtained by dressing rubidium 87 atoms in the $F=1, m_F=-1$ ground state with a radio-frequency field in the presence of a quadrupole static magnetic field of main axis aligned along the vertical axis $z$. The resulting adiabatic trapping potential, within the rotating wave approximation (RWA) for the rf field at frequency $\omega$, writes~\cite{Merloti2013,Garraway2016,Perrin2017}:
\begin{equation}
    V(\bm{r})=\hbar\sqrt{\delta(\bm{r})^2+\Omega(\bm{r})^2}+Mgz,
    \label{eqn:Vrwa}
\end{equation}
where $\delta(\bm{r})=\omega-\alpha\ell$ is the local detuning, $\alpha$ the horizontal gradient of the magnetic quadrupole field (directions $x$ and $y$) in frequency units, $\ell=\sqrt{x^2+y^2+4z^2}$, $\Omega(\bm{r})$ the local atom-rf field coupling, $M$ the atomic mass and $g$ the gravitational acceleration. The potential confines the atoms near the resonant ellipsoidal surface, a shell defined by $\delta(\bm{r})=0$ or equivalently $\ell=r_0$ with $r_0=\omega/\alpha$. For the choice of a circular rf polarization $\sigma^-$ of main axis $z$, $\Omega(\bm{r})=\Omega_0/2\times(1-2z/\ell)$, with $\Omega_0$ the maximum coupling obtained at the bottom of the resonant ellipsoid. In Sec.~\ref{sec:model} we discuss how Eq.~\eqref{eqn:Vrwa} is modified by beyond-RWA effects, but it is sufficient for a qualitative discussion of our results. It is important to note that the potential depends on three parameters $\omega$, $\Omega_0$ and $\alpha$ that are precisely controlled \cite{Note1}, see Appendix \ref{app:calibrations}.

For $\alpha$ below a critical value \cite{Guo2022}, the potential minimum of Eq.~\eqref{eqn:Vrwa} sits at the bottom of the shell, slightly shifted below the resonant ellipsoid $\ell=r_0$ by gravity~\cite{Merloti2013}. Near the minimum, the trap potential is well approximated by a harmonic potential, with radial frequency
\begin{equation}
\omega_{r, \textrm{RWA}}=\sqrt{\frac{g}{4R}}\left(1-\frac{\hbar\Omega_0}{2MgR}\sqrt{1-\epsilon^2}\right)^{1/2}
\label{eqn:freqRWA}
\end{equation}
where $\epsilon=Mg/2\hbar\alpha$ is the gravity sag and
\begin{equation}
R=\frac{r_0}{2}\left(1+\frac{\epsilon}{\sqrt{1-\epsilon^2}}\frac{\Omega_0}{\omega}\right)
\label{eqn:equilibrium_position}
\end{equation}
gives the equilibrium position $z_{\rm eq}=-R$. Eq.~\eqref{eqn:freqRWA} can be interpreted as the oscillation frequency of a rigid pendulum of length $4R$ in an effective reduced gravitational field 
\begin{equation}
g_{\rm eff} = \left(1-\frac{\hbar\Omega_0}{2MgR}\sqrt{1-\epsilon^2}\right)g.
%= \left(1-\frac{\Omega_0}{2\epsilon\omega}\frac{1-\epsilon^2}{\sqrt{1-\epsilon^2}+\epsilon\frac{\Omega_0}{\omega}}\right)g
\end{equation}
However if the atoms are set into motion they explore the curved surface and anharmonic corrections become important~\cite{Guo2020,Sharma2024}. In the following, we evidence this effect as we excite the center-of-mass motion of the quantum gas.

\section{Experimental results}
\label{sec:exp}
Our experimental setup is described in~\cite{Merloti2013,Rey2022}. In short we produce a Bose-Einstein condensate of $^{87}$Rb atoms polarized in the $F=1, m_F=-1$ ground state in a hybrid optical and magnetic trap. We then transfer the quantum gas into the rf-dressed quadrupole trap described in Sec.~\ref{sec:bubble} \cite{Rey2022}.

\begin{figure}[t]
    \centering
    \includegraphics[width=8.6cm]{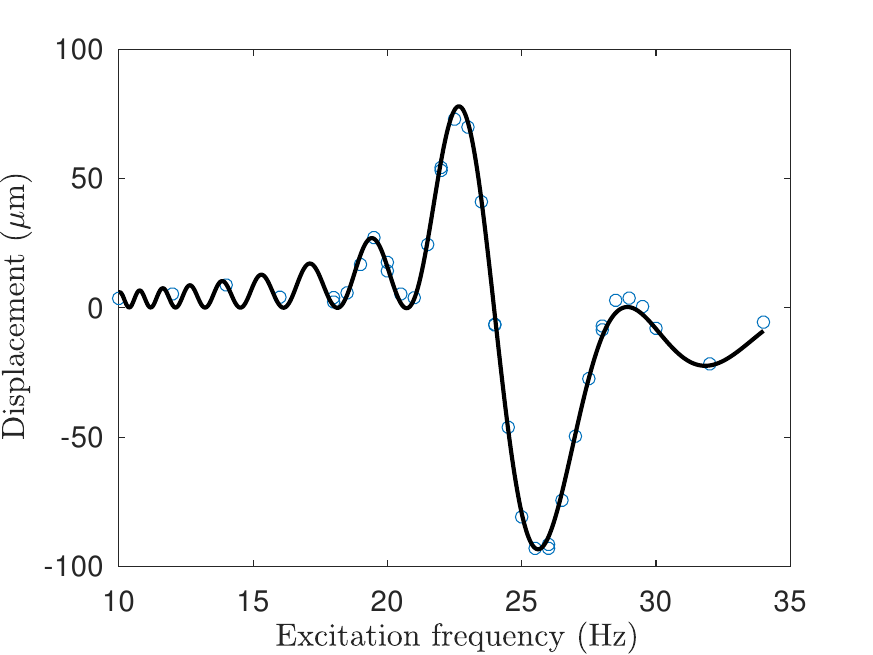}
    \caption{(color online) Measured horizontal position of the cloud after a \SI{23}{ms} time-of-flight as a function of the applied excitation frequency (open blue circles). The black curve is a fit to the data using the driven harmonic oscillator model Eq.~\eqref{eqn:harmonic}, see text for details.
    \label{fig:resonance}
    }
\end{figure}

We perform a resonant excitation spectroscopy of the center-of-mass oscillations of the cloud along a horizontal direction $x$ by applying a uniform modulated magnetic field. This field globally shifts the whole shell and thus the trap position along $x$ by an amount $d\sin\omega_d t$ where $d$ is calibrated independently, see Appendix~\ref{app:calibrations}, resulting in a harmonic drive. We excite the cloud for a duration $6\times 2\pi/\omega_d$ corresponding to 6 periods of the drive. We then abruptly switch off the trapping potential and let the atoms expand in free fall for \SI{23}{ms} before recording a picture of the cloud by resonant absorption imaging along the $y$ axis. We measure the center-of-mass coordinate of the atomic density distribution along $x$ and study its variation as a function of the excitation frequency, see Fig.~\ref{fig:resonance}. A clear resonant behavior is observed, with a global shape of a derivative of a Lorentzian, which allows a very sensitive determination of the central frequency.

We model the center-of-mass motion by a driven harmonic oscillator centered at $x_0$, of frequency $\omega_r$ and damping rate $\gamma$, driven at frequency $\omega_d$ with an amplitude $d$:
\begin{equation}
\ddot{x}+\gamma\dot{x}+\omega_r^2(x-x_0)=d\omega_r^2\sin{\omega_d t}.
%\ddot{x}+\gamma\dot{x}+\omega_r^2x=a\sin{\omega_d t},
\label{eqn:harmonic}
\end{equation}
This equation can be solved analytically, assuming that the cloud is initially at rest, see Appendix~\ref{app:resonance_model}. The position of the cloud after an excitation time $T$ and a free fall of duration $\tau$ is $x_{\rm tof}=x(T)+\tau \dot{x}(T)$. 

We compare this model to the experimental measurement, see Fig.~\ref{fig:resonance}. All the features are captured by the driven harmonic oscillator model. We fit the model to the experimental data and extract $x_0$, $\omega_r$ and $\gamma$. We also fit the excitation amplitude $d_{\rm fit}$, to be compared to the applied calibrated displacement $d$ for a self-consistency check, see Sec.~\ref{sec:diss}.

\begin{figure}
    \centering
    \includegraphics[width=8.6cm]{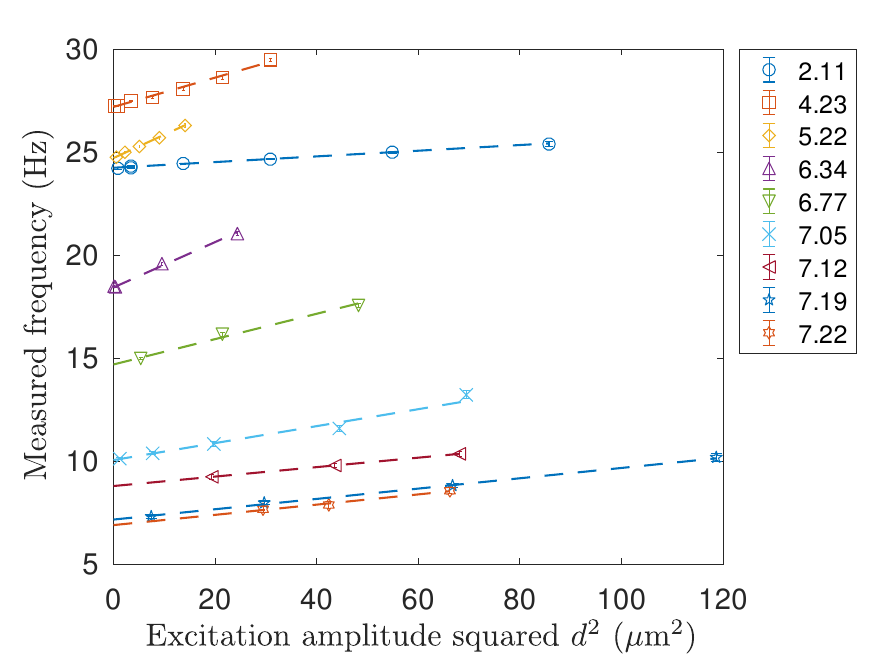}
    \caption{(color online) Radial oscillation frequency deduced from a fit of the data with the driven harmonic oscillator model Eq.~\eqref{eqn:harmonic}, as a function of the excitation amplitude squared $d^2$ (symbols). Each symbol corresponds to a different value of the magnetic gradient $\alpha/(2\pi)$, given in units of \SI{}{\kilo\hertz\per\micro\metre} in the legend. The colored dashed lines are linear fits of each series, allowing to extrapolate the value of the harmonic frequency at vanishing excitation amplitude.
    \label{fig:extrapolation}
    }
\end{figure}

For a purely harmonic trap, the measured frequency should not depend on the excitation amplitude. However, we repeat the measurement for various excitation amplitudes and observe that the fitted values do depend quadratically on the excitation amplitude $d$, see Fig.~\ref{fig:extrapolation}. This an indication that the simple driven harmonic model is only approximate and that beyond harmonic corrections are relevant. We discuss this point in more detail in Sec.~\ref{sec:diss}. Nevertheless, an accurate determination of the resonant frequency is obtained from a straightforward extrapolation of the resonant frequency for vanishing excitation amplitude, see Fig.~\ref{fig:extrapolation}.

We repeat this protocol for increasing values of the magnetic field gradient $\alpha$ and report the values in Fig.~\ref{fig:1} in order to identify precisely the critical gradient that compensates the gravitational acceleration, yielding $g_{\rm eff}=0$. The experimental data (blue circles) are compared to the prediction of Eq.~\eqref{eqn:freqRWA} (yellow dashed line), evaluated using an independent measurement of the trap parameters $\omega$, $\Omega_0$ and $\alpha$, see Appendix~\ref{app:calibrations}.

We observe that while the model reproduces the general behavior of $\omega_r$, the agreement is not quantitative, and the RWA prediction and the measurements differ significantly as the gradient increases. This is due to the rather large value of the Rabi frequency as compared to the rf frequency, $\Omega_0 \simeq 0.29\,\omega$, which requires the addition of beyond-RWA terms in the model.

\section{Corrections beyond the rotating wave approximation}
\label{sec:model}
To improve the trap potential description beyond the rotation wave approximation of Eq.~\eqref{eqn:Vrwa}, we use standard methods, well documented in the literature \cite{Hofferberth2007}. Since the derivation is quite technical, we provide here a summary with the main ideas, while a detailed derivation of the results is given in Appendix~\ref{app:Floquet}.

We use a systematic Floquet expansion to treat beyond-RWA terms, resulting in an infinite set of equations, describing the coupled manifolds. We truncate this expansion at a given order and diagonalize the effective Hamiltonian to extract the resulting potential. We also derive an approximate solution, valid at leading order in $\Omega_0/\omega$, that enables improved analytical predictions for the dressed quadrupole trap geometry. 

\begin{figure}
    \centering
    \includegraphics[width=8.6cm]{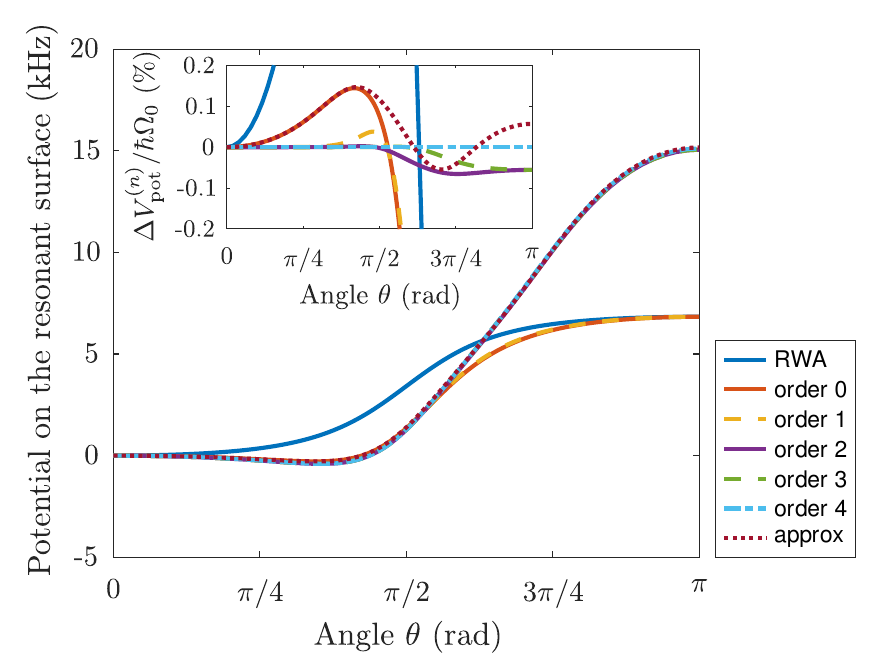}
    \caption{(color online) Value of the potential on the resonance surface as a function of the polar angle $\theta$, measured from the bottom of the ellipsoid, evaluated up to various orders of the Floquet expansion, see text for details. Parameters: $\Omega_0/2\pi=\SI{100}{kHz}$, $\omega/2\pi=\SI{300}{kHz}$, $\alpha/2\pi=\SI{6.0}{\kilo\hertz\per\micro\metre}$. Inset: difference between the potential estimated up to order $n$ and the one estimated up to fourth order, relative to $\hbar\Omega_0$, showing that the lower hemisphere $\theta\in[0,\pi/2]$ is already well captured with an estimation up to second order. The potential given by the analytic expression Eq.~\eqref{eqn:bRWA} is also shown for comparison, dotted red line.
    \label{fig:convergence}
    }
\end{figure}

To test the convergence of the Floquet expansion and the accuracy of the approximate method, we compare the predictions for the potential on the resonance surface $\ell=r_0$ for different truncation orders, as shown in Fig.~\ref{fig:convergence}. This choice simplifies the analysis as, for our rotationally invariant geometry, the resulting potential depends only on the polar angle $\theta$ related to the latitude on the spheroidal resonant surface ($\theta=0$ at the bottom). Moreover this choice ensures that the convergence depends only on the ratio $\Omega_0/\omega$ and not on the value of $\alpha$. Fig.~\ref{fig:convergence} shows that for a ratio $\Omega_0/\omega=1/3$ the Floquet expansion converges reasonably well already at the second order.

In Appendix~\ref{app:Floquet} we derive an improved analytical formula for the trap potential that takes the form:
\begin{equation}
V_{\textrm{Fl}}(\bm{r})=\hbar\sqrt{\left[\delta(\bm{r})-\Sigma(\bm{r})\right]^2+\tilde{\Omega}(\bm{r})^2}+Mgz,
\label{eqn:bRWA}
\end{equation}
where $\Sigma(\bm{r})$ is analogous to a light shift induced by off-resonant coupling terms, mainly due to the local $\sigma^+$ component of the rf field, and $\tilde{\Omega}(\bm{r})$ is the effective coupling due to the local $\sigma^-$ component, renormalized by the local $\pi$ component of the rf field. Explicit expressions of $\Sigma$ and $\tilde{\Omega}$ are given in Appendix~\ref{app:Floquet}. Equation~\eqref{eqn:bRWA} is reminiscent of the one describing the adiabatic potential for an atom interacting simultaneously with several rf fields with well separated frequencies~\cite{Courteille2006}, for which off-resonant fields induce similar light shifts. 
As shown in Fig.~\ref{fig:convergence} this approximate solution is accurate within less than a percent.

Expanding the potential of Eq.~\eqref{eqn:bRWA} around the equilibrium position, we obtain a corrected estimate for the radial oscillation frequency, beyond RWA:
\begin{equation}
    \omega_{r,\textrm{Fl}}=\sqrt{\frac{g}{4R}}\left[1-\frac{\hbar\Omega_0}{2MgR}\sqrt{1-\epsilon^2}\left(1+\frac{\Omega_0^2}{\omega^2}\right)\right]^{1/2}.
    \label{eqn:omega_r}
\end{equation}
This expression is plotted as a red full line in Fig.~\ref{fig:1}. It agrees well with the measurements, and largely improves the RWA prediction.
Importantly Eq.~\eqref{eqn:omega_r} provides an accurate analytical prediction of the condition for gravity compensation in an rf dressed quadrupole trap, without the need of a time-consuming numerical simulation~\cite{Guo2022}.

\section{Discussion}
\label{sec:diss}

The measurements of Fig.~\ref{fig:extrapolation} show that the measured frequency depends quadratically on the excitation amplitude $d$. This is reminiscent of the behavior of an anharmonic oscillator where the frequency depends on the oscillation amplitude. The simplest approximation describing the motion of the atoms in our trap is a rigid pendulum on an ellipsoid : $\bm{r}=r(\theta)\bm{e}_r$, where $r(\theta)=r_0/\sqrt{1+3\cos^2{\theta}}$ and $\theta$ is the polar angle of spherical coordinates. From the classical equation of motion between the turning points $\pm\theta_0$ we obtain:
\[
T \simeq 2\pi\sqrt{\frac{2r_0}{g}}\left(1-\frac{\theta_0^2}{32}\right)
\]
at lowest order in $\theta_0^2$, which shows that the oscillation frequency increases quadratically with the oscillation amplitude, in qualitative agreement with the measurements presented in Fig.~\ref{fig:extrapolation}. This also justifies the use of a harmonic plus quartic potential to describe the atomic motion close to the bottom of the ellipsoid~\cite{Guo2022,Sharma2024}.

\begin{figure}
    \centering
    \includegraphics[width=8.6cm]{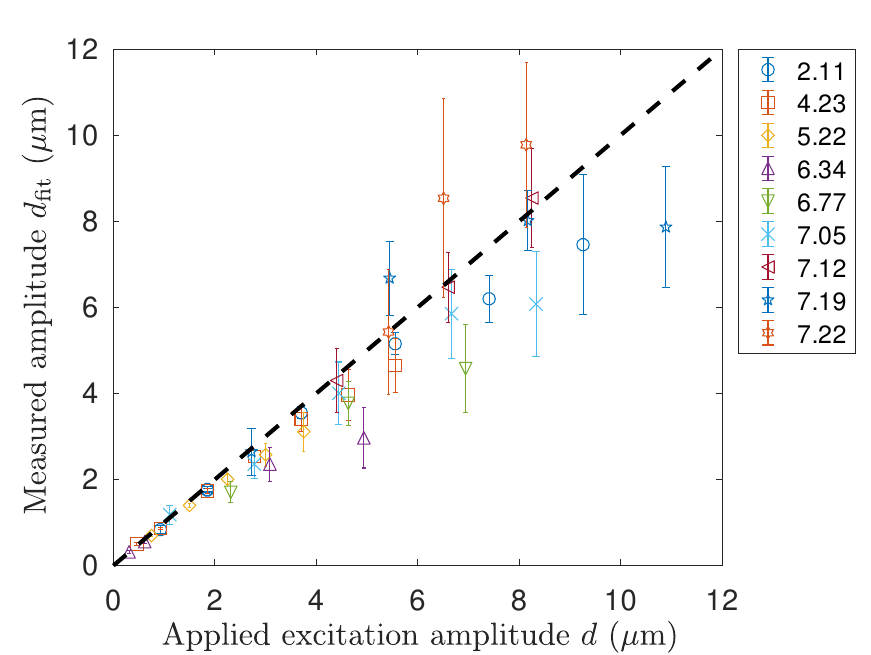}
    \caption{(color online) Measured excitation amplitude $d_{\rm fit}$ as a function of the applied excitation amplitude $d$ (symbols), for different trap gradients (same color code as in Fig.~\ref{fig:extrapolation}). The dashed line corresponds to the identity function and serves as a guide to the eye.
    \label{fig:amplitude}
    }
\end{figure}

One could wonder if the simple harmonic approximation of Eq.~\eqref{eqn:harmonic} is relevant to describe accurately the motion on the curved surface. Figure~\ref{fig:amplitude} shows the fitted excitation amplitude $d_{\rm fit}$, inferred from the fit of resonance curves with the model Eq.~\eqref{eqn:harmonic}, as a function of the applied excitation amplitude $d$ calibrated independently, see Appendix~\ref{app:calibrations}. In the limit of small excitation amplitude the two quantities are indeed equal, showing that the simple driven harmonic model is self consistent. Therefore we believe that our method to extrapolate the bare oscillation frequency is robust and does not depend on the choice of model for the resonance shape, at least for vanishing oscillation amplitude.

\begin{figure}
    \centering
    \includegraphics[width=8.6cm]{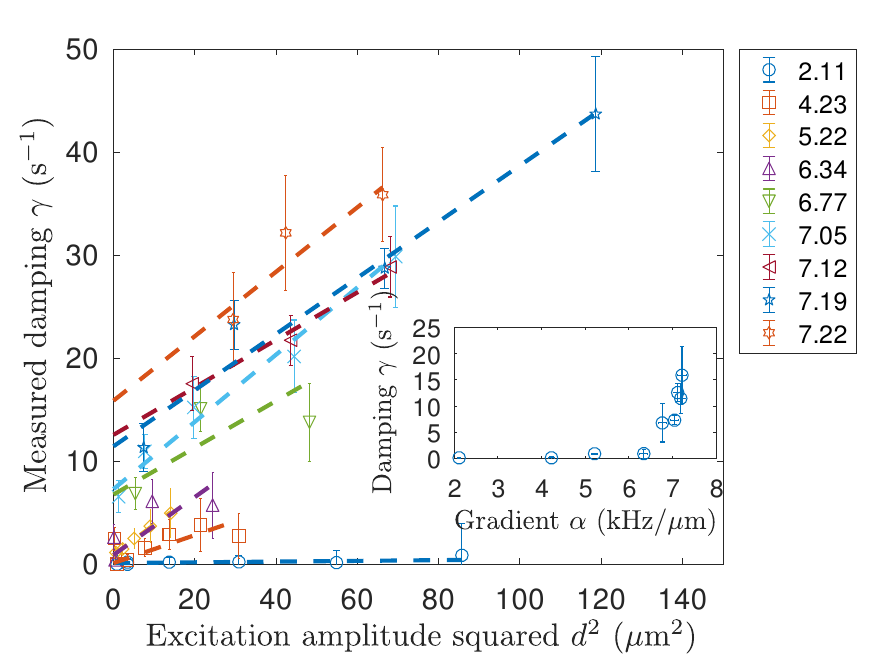}
    \caption{(color online) Measured damping rate as a function of the square of the excitation amplitude (open symbols) for different trap gradients (same color code as in Fig.~\ref{fig:extrapolation}). The black dashed lines are linear fits of the data, allowing to extrapolate the value of the damping rate for vanishing excitation amplitude. Inset: extrapolated damping rate as a function of the trap gradient.
    \label{fig:damping}
    }
\end{figure}

Our model includes an empirical damping term, parameterized by the damping rate $\gamma$, that is necessary to account for the resonances shapes, but cannot be justified at the single particle level: since the potential is conservative, the motion should not be damped and the energy should be conserved. Figure~\ref{fig:damping} shows that for all values of the magnetic field gradient the damping rate tends to increase with the excitation amplitude. This behavior is approximately quadratic and we use a fit to extrapolate the bare damping rate for a vanishing oscillation amplitude, represented in the inset of Fig.~\ref{fig:damping}. We observe a sharp increase of the bare damping rate for gradients larger than \SI{6.7}{\kilo\hertz\per\micro\metre}. We attribute this to a coupling of the center-of-mass motion to the collective modes of the Bose-Einstein condensate, mediated by the trap anharmonicity~\cite{Tanghe2026}. This interpretation is consistent with the fact that when we increase the gradient, the radius of the resonant ellipsoid shrinks and simultaneously the condensate expands on a larger fraction of the curved surface~\cite{Guo2022}, such that the description of the system as a point-like oscillator is no longer valid.

Comparing the insets of Fig.~\ref{fig:1}f) and Fig.~\ref{fig:damping}, we notice a clear correlation between the sharp increase of the bare damping rate and the deviation of the measured oscillation frequency with respect to the harmonic approximation. This behavior is explained as follows: for increasing gradients, the \textit{in situ} condensate density profile becomes more homogeneous~\cite{Guo2022} and the center of mass motion couples to sound waves. This is for example what happens in a flat box-like potential~\cite{Navon2016,Garratt2019,Galka2022}. In our setup, we may thus expect the resonant frequency to level off at high gradients to a value given by the (inverse) typical time scale for the round trip propagation of sound waves, i.e. the ratio of the speed of sound to twice the system size. We estimate the speed of sound in the sample as $c=\sqrt{g_{\rm int}\bar{n}/M}$ \cite{Zaremba1998,Stringari1998} where $g_{\rm int}$ is the interaction constant and the average atomic density $\bar n$ is computed for $N=2.4\times 10^5$ atoms in the shell potential with $\alpha/(2\pi) = \SI{7.19}{\kilo\hertz\per\micro\metre}$, as described by the Floquet approach up to fourth order, see Appendix~\ref{app:Floquet}. Using the Thomas-Fermi profile and taking into account the curved shape of the cloud, we find a frequency for the round trip of sound waves at the speed of sound of \SI{6.6}{\hertz}. This frequency is represented as a black dashed-dotted line in Fig.~\ref{fig:1}f), and agrees well with the measured resonant frequency at the onset of gravity compensation. Predicting more precisely the crossover from a harmonic point-like oscillation to sound wave propagation would require a full time dependent solution of the three-dimensional Gross-Pitaevskii equation for our trapping parameters, that goes beyond the scope of this work. 

\section{Conclusion}
\label{sec:conc}
In summary, we have presented a careful analysis of the center of mass motion of a quantum gas in a shell-shaped adiabatic potential. Terms beyond the rotating wave approximation need to be taken into account to model accurately the shell potential and estimate the harmonic radial frequency at the bottom of the shell. The atoms undergo a vertical force in the shell that plays the role of an effective gravity field, which can be controlled experimentally, for example with the magnetic field gradient. The excellent agreement between the measured frequencies and the improved model provides a precise determination of the parameters for which this effective gravity acceleration vanishes, even more accurate than the direct observation of the atomic density in the trap \cite{Guo2022}.

Around this point, it is possible to prepare a nearly uniform gas over a large fraction of the curved surface of the ellipsoid \cite{Guo2022}, opening the way to the study of hydrodynamics of quantum gases in a curved geometry. Hydrodynamic effects are already apparent in the resonant excitation frequency for the center of mass motion, which transitions from a harmonic regime to a hydrodynamic regime where the frequency is directly linked to the speed of sound.

As the potential becomes more anharmonic, the damping of the center of mass motion also increases. For a driven system, it means that the excess energy is transferred to internal excitations of the gas \cite{Tanghe2026}. It would be interesting to characterize more precisely the nature, the dynamics and the relaxation of these excitations in the curved geometry, in the spirit of recent experiments exploring the superfluid turbulence in uniform quantum gases \cite{Navon2016,Galka2022,Morris2026}.

\begin{acknowledgments}
LPL is UMR 7538 of CNRS and Université Sorbonne Paris Nord.
This work received financial support from the ANR projects VORTECS (Grant No. ANR-22-CE30-0011) and RELAQS (Grant No. ANR-24-CE30-6525), from the France 2030 `QUTISYM' project ANR-23-PETQ-0002, and from the R\'egion \^Ile-de-France in the framework of DIM SIRTEQ and DIM QuanTiP.
M.C. acknowledges financial support from the Program QuanTEdu-France (Grant No. ANR-22-CMAS-0001 France 2030).
R.D. acknowledges support from the French government under the France 2030 investment plan, as part of the Initiative d'Excellence d'Aix-Marseille Université -- AMIDEX AMX-22-CEI-069.
\end{acknowledgments}

\section*{Data availability}
The data that support the findings of this article are openly
available \cite{Note2}.
\appendix

\section{Calibrations}
\label{app:calibrations}

\begin{figure}[b!]
    \centering
    \includegraphics[width=8.6cm]{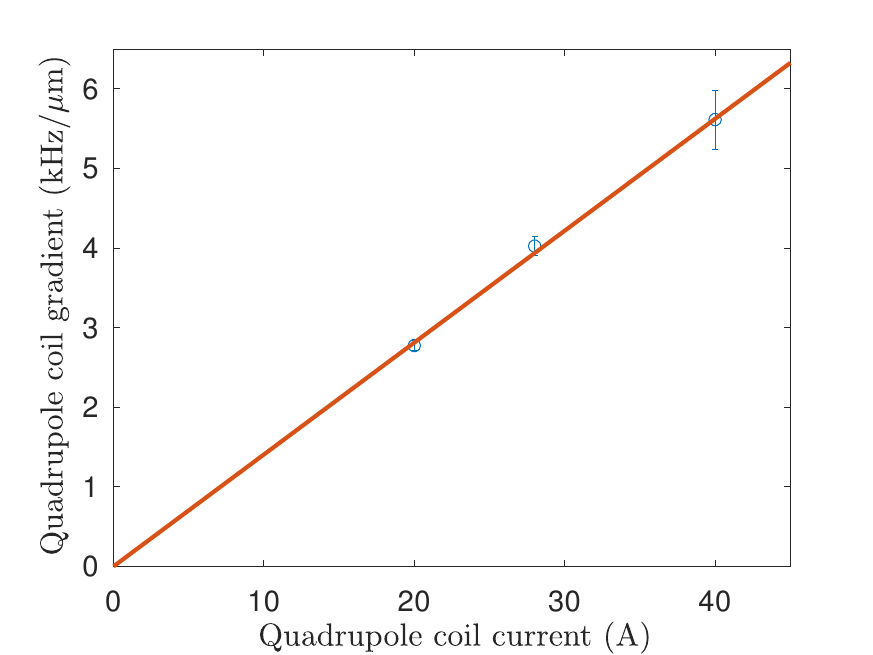}
    \caption{(color online) Calibration of the quadrupole coil horizontal gradient (for reference). Fit function : $y=ax$ with $a=\SI{0.141\pm 0.002}{\kilo\hertz\per\micro\metre\per\ampere}$.
    \label{fig:gradient}
    }
\end{figure}

The quadrupole magnetic field is produced by a pair of water cooled hollow copper conical coils (40 turns each)~\cite{Dubessy2012}. To calibrate the trap gradient $\alpha$ we measure the vertical equilibrium position of the cloud in the dressed potential, deduced from its vertical position after a \SI{23}{ms} free fall, for various values of the dressing frequency between $\SI{300}{kHz}$ and $\SI{900}{kHz}$ and an rf coupling $\Omega_0/2\pi\simeq\SI{44}{kHz}$. For these parameters the gravity sag is small, see Eq.~\eqref{eqn:equilibrium_position}, and the vertical position of the cloud $-R\simeq -r_0/2$ depends linearly on $\omega$. We thus extract the value of the gradient $\alpha$ from a linear fit to the data. We then repeat this process for three different values of the current $I_{\rm quad}$ driving the quadrupole coils, and we find that the gradient is proportional to the current, see Fig.~\ref{fig:gradient}. We can then estimate the gradient for any value of the current: 
$\alpha/(2\pi)=I_{\rm quad}\times\SI{0.141\pm0.002}{\kilo\hertz\per\micro\metre\per\ampere}$.

This measurement depends only on the calibration of the magnification of the imaging system which is performed independently by recording the free fall of a condensate and comparing the measured acceleration to the known value of $g$.

\begin{figure}[b!]
    \centering
    \includegraphics[width=8.6cm]{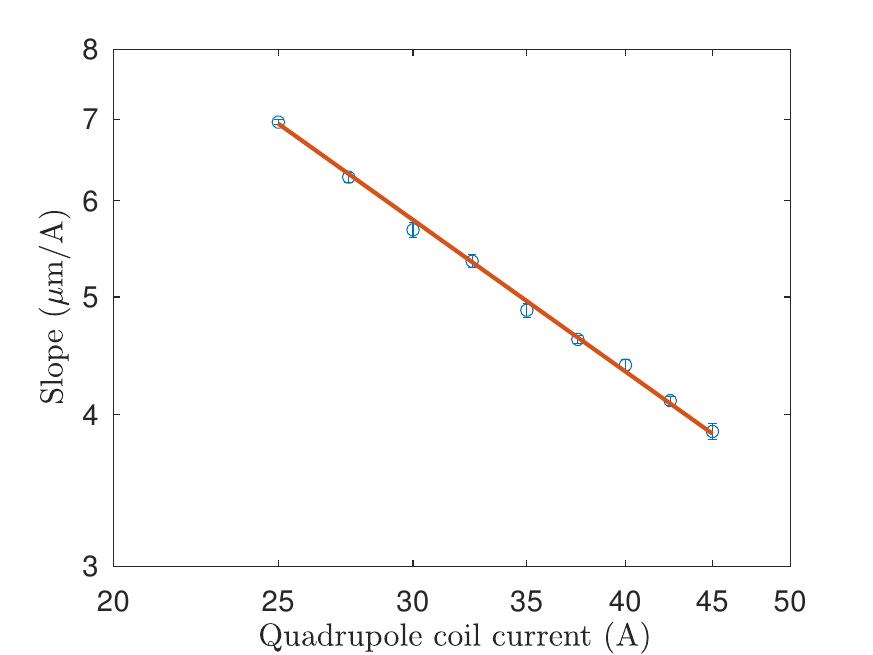}
    \caption{(color online) Calibration of the applied displacement (for reference), in log-log scale. Fit function: $y=b/x$ with $b=\SI{173.6\pm 0.5}{\micro\metre}$.
    \label{fig:displacement}
    }
\end{figure}

The center of the shell trap is shifted when a nearly homogeneous bias field is added to the quadrupole magnetic field. This field is produced by a single excitation coil of axis $x$. The excitation amplitude is calibrated as follows: for a given value $I_{\rm quad}$ of the current in the quadrupole coil, we record the position of the center of mass of the cloud at rest at the bottom of the bubble trap, which varies linearly with the current $I_{\rm exc}$ applied to the excitation coil. The linear slope scales as $1/I_{\rm quad}$, and the final displacement reads $d=b I_{\rm exc}/I_{\rm quad}$ with $b=\SI{173.6\pm 0.5}{\micro\metre}$, see Fig.~\ref{fig:displacement}.

\begin{figure}[b!]
    \centering
    \includegraphics[width=8.6cm]{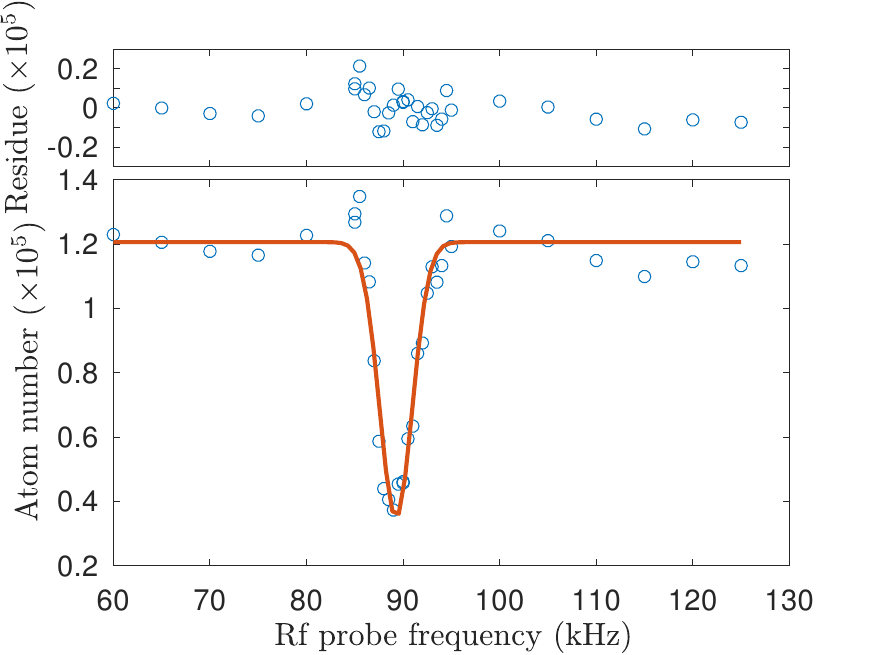}
    \caption{(color online) Calibration of the Rabi coupling. Fit with Gaussian function with $\Omega_{\rm mes}/(2\pi)= \SI{89.26 \pm 0.09}{\kilo\hertz}$.
    Trap parameters: $\omega=2\pi\times\SI{300}{kHz}$, $\alpha/(2\pi)=\SI{4.23\pm0.06}{\kilo\hertz\per\micro\metre}$, circular polarization.
    \label{fig:coupling}
    }
\end{figure}

To calibrate the effective rf-dressing field coupling $\Omega_0$ at the bottom of the shell we perform a rf spectroscopy~\cite{Merloti2013}: we shine a weak rf field at low frequency on the trapped atoms and record the losses as a function of the probe frequency $\omega_{\rm probe}$. When $\omega_{\rm probe}$ is resonant with the $\ket{F=1,m_F=1}\to\ket{F=1,m_F=0}$ transition between dressed states at the trap minimum, it induces losses that we record, see Fig.~\ref{fig:coupling}. Because of the gravity sag the measured resonant frequency is $\Omega_{\rm mes}=\Omega_0/\sqrt{1-\epsilon^2}$. For the gradient $\alpha/(2\pi)=\SI{4.23\pm0.06}{kHz/\micro\metre}$ at which this spectroscopy is realized, we obtain $\epsilon=\SI{0.252\pm0.004}{}$ and $\Omega_0/2\pi=\SI{86.37\pm0.17}{kHz}$.
The rf polarization is finely tuned using three antennas with mutually orthogonal axes, following the protocol described in~\cite{Guo2022}.

\section{Driven harmonic oscillator}
\label{app:resonance_model}
We consider a driven harmonic oscillator in one dimension. The acceleration of the center of mass is given by:
\begin{equation}
\ddot{x}=-\gamma\dot{x}-\omega_r^2(x-x_0-d\sin{\omega_d t}).
%\ddot{x}+\gamma\dot{x}+\omega_r^2x=a\sin{[\omega_d t]}.
\label{eq:forced_damped_harm_osc}
\end{equation}
The aim of this section is to write the explicit solution $x(t)$ for a system initially at rest $x(0)=x_0$ and $\dot{x}(0)=0$, using the standard method to solve second order differential equations with constant coefficients.
Provided that $\omega_r^2>\gamma^2/4$, the homogeneous equation: $\ddot{x}+\gamma\dot{x}+\omega_r^2x=0$ has two linearly independent solutions of the form:
$$
e^{-\gamma t/2}\cos\omega_r't, \qquad e^{-\gamma t/2}\sin\omega_r't,
$$
where $\omega_r'=\sqrt{\omega_r^2-\gamma^2/4}$.

We now need to identify a specific solution of the full equation \eqref{eq:forced_damped_harm_osc}. We look for a solution oscillating at the driving frequency $\omega_d$: $x_0+A\cos\omega_d t + B\sin\omega_d t$. $A$ and $B$ must then verify
\begin{eqnarray*}
-\omega_d^2 A + \gamma\omega_d B + \omega_r^2 A &=& 0\\
-\omega_d^2 B - \gamma\omega_d A + \omega_r^2 B &=& a
\end{eqnarray*}
where $a=M\omega_d^2$ is the acceleration from the drive. Hence
\[
A = -\frac{a\gamma\omega_d}{(\omega_d^2-\omega_r^2)^2+\gamma^2\omega_d^2}~\textrm{and}~B = -\frac{a(\omega_d^2-\omega_r^2)}{(\omega_d^2-\omega_r^2)^2+\gamma^2\omega_d^2}.
\]
The general solution is the sum of this specific solution and the solution of the homogeneous equation $\left(\lambda \cos\omega_r't + \mu \sin\omega_r't\right)e^{-\gamma t/2}$, where $\lambda$ and $\mu$ are set by the initial conditions $x(0)=x_0$ and $\dot x(0)=0$. We get
\[
\lambda = a\frac{\gamma\omega_d}{(\omega_d^2-\omega_r^2)^2+\gamma^2\omega_d^2}
\]
and
\[
\mu = a\frac{\omega_d}{\omega_r'}\frac{\omega_d^2-\omega_r^2+\gamma^2/2}{(\omega_d^2-\omega_r^2)^2+\gamma^2\omega_d^2}.
\]

We finally find:
\begin{widetext}
\begin{equation}
x(t) = x_0+\frac{a}{(\omega_d^2-\omega_r^2)^2+\gamma^2\omega_d^2}\left\{\left[\gamma\omega_d\cos\omega_r't + \frac{\omega_d}{\omega_r'}\left(\omega_d^2-\omega_r^2+\frac{\gamma^2}{2}\right)\sin\omega_r't\right]e^{-\frac{\gamma t}{2}} - \gamma\omega_d\cos\omega_d t - (\omega_d^2-\omega_r^2)\sin\omega_d t\right\}.
\label{eqn:pos}
\end{equation}
\end{widetext}
In the experiment, we have access to the position $x_{\rm tof}(t)$ after a time-of-flight expansion of duration $\tau$, which is related to the velocity $\dot x(t)$ before expansion through $x_{\rm tof}(t) = x(t)+\tau \dot x(t)$. The time derivative of the position can be evaluated explicitly:
\begin{widetext}
\begin{equation}
\dot x(t) = \frac{a\omega_d}{(\omega_d^2-\omega_r^2)^2+\gamma^2\omega_d^2}\left\{\left[-\frac{\gamma}{2\omega_r'}\left(\omega_d^2+\omega_r^2\right)\sin\omega_r't + (\omega_d^2-\omega_r^2)\cos\omega_r't\right]e^{-\frac{\gamma t}{2}} + \gamma\omega_d\sin\omega_d t - (\omega_d^2-\omega_r^2)\cos\omega_d t \right\}.
\label{eqn:vel}
\end{equation}
\end{widetext}
We now turn to the measurement done in the experiment, where we record the value of $x_{\rm tof}\left(t\right)$ after $n$ periods, such that $t=2\pi n/\omega_d$, as a function of 
$\omega_d$. 
This choice ensures that $\sin\omega_d t=0$ and $\cos\omega_d t=1$.
Simplifying Eqs.~\eqref{eqn:pos} and \eqref{eqn:vel} we obtain an analytical formula that we use to fit our measurements. Although straightforward to derive this formula is quite long and we do not report it here. We stress however that both terms $x(t)$ and $\dot x(t)$ are important to predict accurately the shape of the resonance, as shown in Fig.~\ref{fig:resonance}.

\clearpage
\section{Floquet expansion for a dressed trap}
\label{app:Floquet}
In this section we present a systematic method to take into account the atom-rf field coupling in a dressed trap configuration, including beyond-RWA corrections. We show that the exact result can be approximated by an improved analytical formula, where the physical origin of the corrections are evidenced.

\subsection{Method}
The atom-field Hamiltonian reads:
\begin{equation}
%    \hat{H}=\frac{\hat{\bm{p}}^2}{2M}+\frac{\mu_B g_F}{\hbar}\hat{\bm{F}}\cdot\left[\bm{B}_0(\bm{r})+\bm{B}(\bm{r},t)\right],
    %\hat{H}=\frac{\hat{\bm{p}}^2}{2M}+\frac{g_F\mu_B}{\hbar}\left[\bm{B}_0(\bm{r})+\bm{B}(\bm{r},t)\right]\cdot\hat{\bm{F}},
    \hat{H}=\frac{g_F\mu_B}{\hbar}\left[\bm{B}_0(\bm{r})+\bm{B}(\bm{r},t)\right]\cdot\hat{\bm{F}},
    \label{eqn:HC1}
\end{equation}
where $\bm{B}_0(\bm{r})$ is the static magnetic field and $\bm{B}(\bm{r},t)$ is the rf magnetic field, oscillating at frequency $\omega$ \cite{Note3}.
In Eq.~\eqref{eqn:HC1} we omitted the kinetic energy part of the Hamiltonian because in the following we always assume that the adiabatic approximation holds, such that it amounts only to an energy shift.
The first step towards the solution~\cite{Garraway2016,Perrin2017} is to apply a spin rotation to align the quantization axis with the local magnetic field orientation, resulting in:
\begin{equation}
%    \hat{H}_0\simeq\frac{\hat{\bm{p}}^2}{2M}+\romain{s}\omega_0(\bm{r})\hat{F}_z+\frac{\mu_B g_F}{\hbar}\hat{\bm{F}}\cdot\tilde{\bm{B}}(\bm{r},t),
    \hat{H}_0\simeq s\omega_0(\bm{r})\hat{F}_z+\frac{g_F\mu_B}{\hbar}\tilde{\bm{B}}(\bm{r},t)\cdot\hat{\bm{F}},
    \label{eqn:spin_rotation}
\end{equation}
where $\omega_0(\bm{r})=|g_F|\mu_B|\bm{B}_0(\bm{r})|/\hbar>0$ is the local Larmor frequency, $s=g_F/|g_F|=\pm1$ is the sign of $g_F$, and  $\tilde{\bm{B}}(\bm{r},t)$ is the rf field in the new frame. Detailed formula applied to the rf dressed quadrupole trap are given in Sec.~\ref{sec:appBRWA}.

It is useful to write Eq.~\eqref{eqn:spin_rotation} in the spherical basis as:
\begin{eqnarray}
\nonumber
    \hat{H}_0&\simeq&s\left\{\left[\omega_0(\bm{r})+\Omega_z(\bm{r},t)\right]\hat{F}_z
    \right.\\
    &&\left.+\frac{\Omega_+(\bm{r},t)}{2}\hat{F}_++\frac{\Omega_-(\bm{r},t)}{2}\hat{F}_-\right\},
    \label{eqn:spin_rotation2}
\end{eqnarray}
where $\hat{F}_\pm=\hat{F}_x\pm i\hat{F}_y$ are the spin raising and lowering operators respectively and 
\begin{eqnarray*}
\Omega_z(\bm{r},t)&=&\frac{|g_F|\mu_B}{\hbar}\tilde{B}_z(\bm{r},t),\\
\Omega_\pm(\bm{r},t)&=&\frac{|g_F|\mu_B}{\hbar}\left[\tilde{B}_x(\bm{r},t)\mp i\tilde{B}_y(\bm{r},t)\right],
\end{eqnarray*}
are the time dependent atom-field couplings, with $\tilde{B}_i(\bm{r},t)=\tilde{\bm{B}}(\bm{r},t)\cdot\mathbf{e}_i$ the rf field projections on the axes in the new frame. We remark that $\Omega_-(\bm{r},t)=[\Omega_+(\bm{r},t)]^*$. In contrast to what is usually done we will not assume a specific rf field polarization and hence keep the contributions coming from all three couplings. This is motivated by the fact that when the atoms move on the surface of the bubble trap, they experience different rf polarization depending on their position.

We now apply a spin rotation $\hat{R}=\exp{[-is(\omega t+f(\bm{r},t))\hat{F}_z/\hbar]}$ to remove the time dependence of the $\hat{F}_z$ term, where $f(\bm{r},t)=\int^t dt^\prime\,\Omega_z(\bm{r},t^\prime)$. The transformed Hamiltonian $\hat{H}_1=\hat{R}^\dagger\hat{H}_0\hat{R}$ reads:
\begin{eqnarray}
\nonumber
    \hat{H}_1&\simeq&-s\delta(\bm{r})\hat{F}_z
    +s\frac{\Omega_+(\bm{r},t)}{2}e^{s i(\omega t+f(\bm{r},t))}\hat{F}_+\\
    &&+s\frac{\Omega_-(\bm{r},t)}{2}e^{-s i(\omega t+f(\bm{r},t))}\hat{F}_-,
    \label{eqn:H1}
\end{eqnarray}
where $\delta(\bm{r})=\omega-\omega_0(\bm{r})$ is the local detuning.

Since $\bm{B}(\bm{r},t)$ has no DC component, $f(\bm{r},t)$ is periodic in time with the same period as the rf field $T=2\pi/\omega$ and we may introduce a formal Fourier series expansion: $e^{isf(\bm{r},t)}=\sum_n c_n(\bm{r})e^{in\omega t}$. Following the Floquet method, we look for a solution of the Schr{\"o}dinger equation under the form: $\ket{\psi}=\sum_{n=-\infty}^\infty e^{i(n\omega t-Et/\hbar)}\ket{\psi_n}$, where the basis is orthogonal such that $\braket{\psi_n|\psi_{n^\prime}}=0$ for $n\neq n'$ and the normalization of the wave function imposes $\sum_n\braket{\psi_n|\psi_n}=1$.

The eigenvalue equation $E\ket{\psi}=\hat{H}_1\ket{\psi}$ results in an infinite set of coupled equations:
\begin{equation}
    E\ket{\psi_n}\simeq\hat{D}_n(\bm{r})\ket{\psi_n}+\sum_{k\neq0}\hat{V}_k(\bm{r})\ket{\psi_{n+k}},
    \label{eqn:Floquet}
\end{equation}
where, again we have neglected the spin-orbit coupling terms. In Eq.~\eqref{eqn:Floquet} the diagonal term is:
\[
\hat{D}_n(\bm{r})=n\hbar\omega\hat{I}-s\delta(\bm{r})\hat{F}_z+\hat{V}_0(\bm{r}).
\]
Here, $\hat{V}_0(\bm{r})=\frac{s}{2}\sum_l c_l(\bm{r}) \tilde{\Omega}_+^{(l+s)}(\bm{r})\hat{F}_+ + h.c.$ is hermitien
and the off-diagonal couplings, verifying $\hat{V}_{-k}=\hat{V}_k^\dagger$, are:
\[
\hat{V}_k(\bm{r})=\frac{s}{2}\sum_l\left[c_l(\bm{r})\tilde{\Omega}_+^{(l+s+k)}(\bm{r})\hat{F}_+
+c_l^*(\bm{r})\tilde{\Omega}_-^{(l+s-k)}(\bm{r})\hat{F}_-\right],
\]
where $\tilde{\Omega}_\pm^{(q)}(\bm{r})=1/T\times\int_0^T dt\,\Omega_\pm(\bm{r},t)e^{\pm iq\omega t}=[\tilde{\Omega}_\mp^{(q)}(\bm{r})]^*$ are the $q$-th coefficients of the Fourier series. We note that for a single-frequency rf field only the $q=\pm1$ harmonics contribute, and that we recover the usual RWA approximation in the limit $c_l(\bm{r})\to\delta_{l,0}$, keeping only the $\hat{D}_n(\bm{r})$ contribution in Eq.~\eqref{eqn:Floquet}.

Equation~\eqref{eqn:Floquet} can be used to compute the adiabatic potential with arbitrary precision by truncating the expansion at a given order. For example, at first order one has to solve the eigenvalue problem:
\[
E\begin{pmatrix}\ket{\psi_1}\\\ket{\psi_0}\\\ket{\psi_{-1}}\end{pmatrix}
=\begin{pmatrix}
\hat{D}_0+\hbar\omega\hat{I}&\hat{V}_{-1}&\hat{V}_{-2}\\
\hat{V}_1&\hat{D}_0&\hat{V}_{-1}\\
\hat{V}_{2}&\hat{V}_{1}&\hat{D}_0-\hbar\omega\hat{I}
\end{pmatrix}\begin{pmatrix}\ket{\psi_1}\\\ket{\psi_0}\\\ket{\psi_{-1}}\end{pmatrix}.
\]
Finally, to compute the effective potential for the slow atomic motion, we need to obtain the quasi-energies of the central ($n=0$) manifold.

\subsection{Formal solution}
Equation~\eqref{eqn:Floquet} can be formally solved using operator algebra~\cite{Cohen1998} and at leading order in $1/\omega$ we obtain:
\[
E\ket{\psi_0}\simeq\left(-s\delta(\bm{r})\hat{F}_z+\hat{V}_0(\bm{r})+\sum_{n>0}\frac{\left[\hat{V}_n^\dagger,\hat{V}_n\right]}{n\hbar\omega}\right)\ket{\psi_0}.
\]
It is straightforward to show that $[\hat{V}_n^\dagger,\hat{V}_n]$ is always proportional to $\hat{F}_z$ and $\hat{V}_0(\bm{r})=\frac{s}{2}\left[\tilde{\Omega}_+(\bm{r})\hat{F}_++\tilde{\Omega}_-(\bm{r})\hat{F}_-\right]$, where $\tilde{\Omega}_+(\bm{r})=[\tilde{\Omega}_-(\bm{r})]^*=\sum_lc_l(\bm{r})\tilde{\Omega}_+^{(l+s)}(\bm{r})$. 
We introduce the modulus $\tilde{\Omega}=|\tilde{\Omega}_+|$ and phase $\varphi$ of $\tilde{\Omega}_+$ and write
\[
s \tilde{\Omega}_+(\bm{r}) = \tilde{\Omega}(\bm{r}) e^{-i\varphi(\bm{r})}.
\]
Therefore the effective Hamiltonian is:
\[
%\hat{H}=\frac{\hat{\bm{p}}^2}{2M}-\romain{s}(\delta(\bm{r})-\Sigma(\bm{r}))\hat{F}_z+\romain{s}\Ln{}{\frac{\tilde{\Omega}_+(\bm{r})}{2}}\hat{F}_++\romain{s}\Ln{}{\frac{\tilde{\Omega}_-(\bm{r})}{2}}\hat{F}_-,
\hat{H}=-s(\delta(\bm{r})-\Sigma(\bm{r}))\hat{F}_z
+\frac{\tilde{\Omega}(\bm{r})}{2}\left(e^{-i\varphi(\bm{r})}\hat{F}_+ + e^{i\varphi(\bm{r})}\hat{F}_-\right),
\]
where $s\Sigma(\bm{r})\hat{F}_z=\sum_{n>0}[\hat{V}_n^\dagger,\hat{V}_n]/n\hbar\omega$.

The last term can be recast under the form
\[
\tilde{\Omega}(\bm{r})\left[\cos\varphi(\bm{r})F_x + \sin\varphi(\bm{r})F_-\right],
\]
such that the last step is again a spin rotation
\[
\hat{U}_1=\exp{[i\vartheta(\bm{r})(\sin{\varphi(\bm{r})}\hat{F}_x-\cos{\varphi(\bm{r})}\hat{F}_y)/\hbar]},
\]
where the angle $\vartheta$ is defined as
\begin{eqnarray}
\vartheta(\bm{r})&=&\arccos{\frac{-s(\delta(\bm{r})-\Sigma(\bm{r}))}{\sqrt{(\delta(\bm{r})-\Sigma(\bm{r}))^2+\tilde{\Omega}(\bm{r})^2}}},\\ 
\cos{\varphi(\bm{r})}&=&s\frac{\tilde{\Omega}_+(\bm{r})+\tilde{\Omega}_-(\bm{r})}{2\tilde{\Omega}(\bm{r})}, \quad\mbox{and}\\
\sin{\varphi(\bm{r})}&=&is\frac{\tilde{\Omega}_+(\bm{r})-\tilde{\Omega}_-(\bm{r})}{2\tilde{\Omega}(\bm{r})},
\end{eqnarray}
resulting in:
\begin{equation}
%\hat{H}=\frac{\hat{\bm{p}}^2}{2M}+\sqrt{(\delta(\bm{r})-\Sigma(\bm{r}))^2+|\tilde{\Omega}(\bm{r})|^2}\hat{F}_z,
\hat{H}=\sqrt{(\delta(\bm{r})-\Sigma(\bm{r}))^2+\tilde{\Omega}(\bm{r})^2}\hat{F}_z.
\label{eqn:HbRWA}
\end{equation}

Equation~\eqref{eqn:HbRWA} is the main theoretical result, giving the explicit form of the beyond-RWA corrections: the resonant surface condition is now $\delta(\bm{r})=\Sigma(\bm{r})$, where the off-resonant terms contribute to an effective light shift, and the atom-field coupling is renormalized by the $c_l(\bm{r})$ coefficients.

\subsection{Application to the dressed quadrupole trap}
\label{sec:appBRWA}
We now give explicit formula for the geometry of our experiment, for which $s=-1$ and the static field $\bm{B}_0(\bm{r})$ is aligned along the unit vector $\bm{u}(\bm{r})=\frac{1}{\ell}(x\bm{e}_x+y\bm{e}_y-2z\bm{e}_z)$. In the following $(\rho,\phi,z)$ are the cylindrical coordinates and we recall $\ell=\sqrt{\rho^2+4z^2}$. For the sake of simplicity we consider only the case of a uniform rf field with a purely circular $\sigma^-$ rf dressing polarization, and a single dressing frequency:
\[
\bm{B}(t)=B_{\rm rf}\left(\cos{(\omega t)}\bm{e}_x - \sin{(\omega t)}\bm{e}_y\right).
\]
The first spin rotation should bring $\hat{\bm{F}}\cdot\bm{u}$ onto $\hat{F}_z$. This is done by applying the rotation operator $\hat{R}_1=e^{-i\beta \hat{F}_y/\hbar}\,e^{-i\phi \hat{F}_z/\hbar}$ where the angle $\beta$ is defined by $\cos\beta=-2z/\ell$ and $\sin\beta=\rho/\ell$. It yields $\hat{R}_1^\dagger \hat{\bm{F}}\cdot\bm{u} \hat{R}_1 = \hat{F}_z$, and more generally gives on the projections of the spin operator:
\begin{eqnarray*}
\hat{R}_1^\dagger \hat{F}_x\hat{R}_1 = \cos\phi \sin\beta \hat{F}_z + \cos\phi \cos\beta \hat{F}_x - \sin\phi \hat{F}_y,\\
\hat{R}_1^\dagger \hat{F}_y\hat{R}_1 = \sin\phi \sin\beta \hat{F}_z + \sin\phi \cos\beta \hat{F}_x + \cos\phi \hat{F}_y,\\
\hat{R}_1^\dagger \hat{F}_z\hat{R}_1 = \cos\beta \hat{F}_z - \sin\beta \hat{F}_x.
\end{eqnarray*}
This yields for the transformed rf field:
\begin{eqnarray*}
\tilde{B}_x = - B_{\rm rf}\frac{2z}{\ell}\cos(\omega t+\phi)\\
\tilde{B}_y = -B_{\rm rf}\sin(\omega t+\phi),\\
\tilde{B}_z = B_{\rm rf}\frac{\rho}{\ell}\cos(\omega t+\phi),
\end{eqnarray*}
hence
\begin{eqnarray*}
\Omega_\pm(\bm{r},t)&=&\pm\frac{\Omega_0}{2}\left[\left(1\mp\frac{2z}{\ell}\right)e^{i\phi}e^{i\omega t} - \left(1\pm\frac{2z}{\ell}\right)e^{-i\phi}e^{-i\omega t} \right],\\
\Omega_z(\bm{r},t)&=&\Omega_0\frac{\rho}{\ell}\cos{(\omega t+\phi)},
\end{eqnarray*}
where $\Omega_0=|g_F|\mu_BB_{\rm rf}/\hbar$ is the typical coupling strength.
As mentioned above, the atoms interact with the three rf polarization as they move on the resonant surface $\ell=r_0$, because of the spatial structure of the underlying static magnetic field.

We may now compute
\[
f(\bm{r},t)=\frac{\Omega_0}{\omega}\frac{\rho}{\ell}\sin{(\omega t+\phi)},
\]
and using the Jacobi-Anger expansion ($e^{i w\sin{\theta}}=\sum_n J_n[w]e^{in\theta}$), we obtain with $s=-1$:
\[
e^{-if(\bm{r},t)}=\sum_n (-1)^nJ_n\left[\frac{\Omega_0}{\omega}\frac{\rho}{\ell}\right]e^{in(\omega t+\phi)},
\]
where $J_n[w]$ is the $n$-th Bessel function of the first kind and we used $J_n[-w]=(-1)^nJ_n[w]$. We may thus identify: $c_n(\bm{r})=(-1)^nJ_n[\Omega_0\rho/\omega\ell]e^{in\phi}=(-1)^n j_n(\rho) e^{in\phi}$. The only non-vanishing harmonics are:
\[
\tilde{\Omega}_\pm^{(+1)}(\bm{r})=-  \frac{\Omega_0}{2}\left(1 + \frac{2z}{\ell}\right)e^{\mp i\phi}=-  \Omega_+(\bm{r}) e^{\mp i\phi}
\]
and
\[
\tilde{\Omega}_\pm^{(-1)}(\bm{r})=  \frac{\Omega_0}{2}\left(1 - \frac{2z}{\ell}\right) e^{\pm i\phi} =  \Omega_-(\bm{r}) e^{\pm i\phi},
\]
where $\Omega_\pm(\bm{r})=\Omega_0(1\pm2z/\ell)/2$, are the usual amplitudes for the couplings associated to the $\sigma^\pm$ polarizations in the standard RWA treatment of the rf-dressed quadrupole trap~\cite{Garraway2016,Perrin2017}.

Finally the atom-field couplings are:
\begin{eqnarray*}
\hat{V}_k && =\frac{e^{-ik\phi}}{2}\left[-j_k\Omega_-\left(e^{i\phi}\hat{F}_+ + (-1)^k e^{-i\phi}\hat{F}_-\right)\right.\\
&&\left.+\Omega_+\left(j_{k-2}e^{i\phi}\hat{F}_+ + (-1)^k j_{k+2}e^{-i\phi}\hat{F}_-\right)\right],
\end{eqnarray*}
where the position dependence in $\hat{V}_k$, $\Omega_\pm$ and $j_k$ was omitted.

With these expressions it is straightforward to evaluate the terms of Eq.~\eqref{eqn:HbRWA}. For the effective coupling we find:
\[
\tilde{\Omega}(\bm{r})=|j_0(\rho)\Omega_{-}(\bm{r})-j_{2}(\rho)\Omega_{+}(\bm{r})|.
\]
The computation of the light shift is a little bit more involved, we first evaluate:
\[
[\hat{V}^\dagger_k,\hat{V}_k]=\hbar\frac{(j_k\Omega_- - j_{k+2}\Omega_+)^2-(j_k\Omega_- - j_{k-2}\Omega_+)^2}{2}\hat{F}_z
\]
where we used the identity $[\hat{F}_+,\hat{F}_-]=2\hbar\hat{F}_z$. We finally obtain:
\[
\Sigma=\sum_{k>0}\frac{(j_k\Omega_- - j_{k-2}\Omega_+)^2 - (j_k\Omega_- - j_{k+2}\Omega_+)^2}{2k\omega}.
\]
In practice, close to the equilibrium position, at the bottom of the shell potential, the leading order term of this sum will be given by the $k=2$ term with the $j_0$ Bessel function, and we may use:
\[
\Sigma(\bm{r})\simeq\frac{\left[j_{0}(\bm{r})\Omega_+(\bm{r})\right]^2}{4\omega}.
\]

\subsection{Improved harmonic approximation}
In the presence of gravity the total potential is given by Eq.~\eqref{eqn:bRWA}, that we repeat here for convenience:
\[
V_{\textrm{Fl}}(\bm{r})=\hbar\sqrt{[\delta(\bm{r})-\Sigma(\bm{r})]^2+\tilde{\Omega}(\bm{r})^2}+Mgz.
\]
For a gradient below the one required to compensate gravity, the global minimum is at the bottom of the shell potential, at the same position than for the RWA computation $z_{\rm eq}=-R$. This is due to the fact that all Bessel functions $j_{n>0}(\rho)$ vanish on the $z$ axis (and $j_0(\rho)=1$ for $\rho=0$). We then expand the potential in the radial direction around the equilibrium position $\rho=0, z=-R$. The light shift $\Sigma(\bm{r})\simeq O(\rho^4)$ does not contribute to the second order terms, such that the only contribution comes from the reduction of the effective coupling, due to the $\pi$ polarization component. Using $J_0[w]\simeq 1-w^2/4$ for $w\to0$ we get:
\[
\tilde{\Omega}(\bm{r})\simeq-\Omega_0\left[1-\frac{\rho^2}{16R^2}\left(1+\frac{\Omega_0^2}{\omega^2}\right)
\right].
\]
This results in a correction of the coupling strength $\Omega_0$ in the RWA expression of Eq.~\eqref{eqn:freqRWA}, giving immediately the improved formula of Eq.~\eqref{eqn:omega_r}.

%

%\bibliography{biblio}

\end{document}